\documentstyle[12pt,epsf]{article}
\normalbaselines
\begin{document}


\newcommand{\sa}{self-adjoint}
\newcommand{\se}{Schr\"odinger equation}
\newcommand{\bm}{ Bohmian mechanics}
\newcommand{\qf}{quantum formalism}
\newcommand{\qm}{quantum mechanics}
\newcommand{\wf}{wave function}
\newcommand{\ewf}{effective wave function}
\newcommand{\cwf}{conditional wave function}
\newcommand{\oqm}{orthodox \qm}
\newcommand{\qe}{quantum equilibrium}
\newcommand{\rv}{random variable}
\newcommand{\hv}{hidden variable}
\newcommand{\LSE}{Lippmann-Schwinger equation}

\newcommand{\PV}{projection valued measure}
\newcommand{\POV}{positive operator valued measure}

\newcommand{\bd}{\begin{displaymath}}
\newcommand{\ed}{\end{displaymath}}
\newcommand{\be}{\begin{equation}}
\newcommand{\ee}{\end{equation}}
\newcommand{\bi}{\begin{itemize}}
\newcommand{\ei}{\end{itemize}}
\newcommand{\ba}{\begin{eqnarray}}
\newcommand{\ea}{\end{eqnarray}}

\newcommand{\R}{\mbox{${\rm I\!R}$} }
\newcommand{\E}{\mbox{${\rm I\!E}$} }
\newcommand{\ran}{\mbox{\rm ran} }
\newcommand{\supp}{\mbox{\rm supp} }
\newcommand{\F}{{\cal F }}
\newcommand{\PROB}{{\large \bf P}}
\newcommand{\lit}{ \lim_{t \to \infty} }
\newcommand{\slim}{ {\mbox{\rm s-}} \lim}
\newcommand{\lir}{ \lim_{R \to \infty} }
\newcommand{\ai}{ {\alpha\in {\cal I} }}
\newcommand{\psihat}{\hat{\psi}}
\newcommand{\phihat}{\hat{\phi}}
\newcommand{\xt}{\frac{\bx}{t}}
\newcommand{\yt}{\frac{\by}{t}}
\newcommand{\kx}{\bk\cdot\bx}
\newcommand{\xyt}{\frac{\bx\cdot\by}{t}}
\newcommand{\xxt}{\frac{x^2}{2t}}
\newcommand{\yyt}{\frac{y^2}{2t}}
\newcommand{\kkt}{\frac{k^2 t}{2}}
\newcommand{\vvt}{\frac{v^2 t}{2}}
\newcommand{\bj}{{\bf j}}
\newcommand{\bsigma}{{\bf \sigma}}
\newcommand{\bq}{{\bf q}}
\newcommand{\bx}{{\bf x}}
\newcommand{\by}{{\bf y}}
\newcommand{\bz}{{\bf z}}
\newcommand{\bv}{{\bf v}}
\newcommand{\bn}{{\bf n}}
\newcommand{\bk}{{\bf k}}
\newcommand{\bp}{{\bf p}}
\newcommand{\br}{{\bf R}}
\newcommand{\bg}{{\bf g}}
\newcommand{\bnu}{{\bf \bn}}
\newcommand{\rkp}{\widetilde{|\bk'\rangle}}
\newcommand{\lk}{\widetilde{\langle\bk|}}
\newcommand{\rk}{\widetilde{|\bk\rangle}}
\newcommand{\CKR}{{C\cap\PKR}}
\newcommand{\PKR}{{\partial B_R}}
\newcommand{\jpsi}{\bj^{\psi_t}}
\newcommand{\jphi}{\bj^{\phi_t}}
\newcommand{\SIC}{scattering-into-cones theorem}
\newcommand{\FAS}{flux-across-surfaces theorem}
\newcommand{\IM}{\mbox{\rm Im}\, }
\newcommand{\sm}{\setminus}
\newcommand{\proofend} {\rule[0.25ex]{1.0ex}{1.0ex} }
\newtheorem{satz}{Satz}[section]
\newtheorem{theorem}[satz]{Theorem}
\newtheorem{conjecture}[satz]{Conjecture}
\newtheorem{corollary}[satz]{Corollary}
\newtheorem{lemma}[satz]{Lemma}
\newtheorem{definition}[satz]{Definition}
\newtheorem{remark}[satz]{Remark}
\newtheorem{example}[satz]{Example}
\newtheorem{propos}[satz]{Proposition}

\newenvironment{rmdef}{\begin{definition}\rm}{\end{definition}}
\newenvironment{rmex}{\begin{example}\rm}{\end{example}}
\newenvironment{rmrem}{\begin{remark}\rm}{\end{remark}}

\newcommand{\proof}{\paragraph{Proof}}


\def\offen#1{\mathchoice
    { \hbox{${\it \stackrel{o}{#1}}$} }
    { \hbox{${\it \stackrel{o}{#1}}$} }
    { \hbox{$ \scriptstyle  {\it \stackrel{o}{#1}}$} }
    { \hbox{$ \scriptscriptstyle {\it \stackrel{o}{#1}}$} } }
\def\Io{\offen{I}}
\def\Go{\offen{G}}


\def\eitsch{\mathchoice{ \hbox{${\cal H}$} }
                   { \hbox{${\cal H}$} }
                   { \hbox{$ \scriptstyle {\cal H}$} }
                   { \hbox{$ \scriptscriptstyle  {\cal H}$} }   }

\def\IR{\mathchoice{ \hbox{${\rm I}\!{\rm R}$} }
                   { \hbox{${\rm I}\!{\rm R}$} }
                   { \hbox{$ \scriptstyle  {\rm I}\!{\rm R}$} }
                   { \hbox{$ \scriptscriptstyle  {\rm I}\!{\rm R}$} }   }
\def\IP{\mathchoice{ \hbox{${\rm I}\!{\rm P}$} }
                   { \hbox{${\rm I}\!{\rm P}$} }
                   { \hbox{$ \scriptstyle  {\rm I}\!{\rm P}$} }
                   { \hbox{$ \scriptscriptstyle  {\rm I}\!{\rm P}$} }   }
\def\IE{\mathchoice{ \hbox{${\rm I}\!{\rm E}$} }
                   { \hbox{${\rm I}\!{\rm E}$} }
                   { \hbox{$ \scriptstyle  {\rm I}\!{\rm E}$} }
                   { \hbox{$ \scriptscriptstyle  {\rm I}\!{\rm E}$} }   }

\def\IN{\mathchoice{ \hbox{${\rm I}\!{\rm N}$} }
                   { \hbox{${\rm I}\!{\rm N}$} }
                   { \hbox{$ \scriptstyle  {\rm I}\!{\rm N}$} }
                   { \hbox{$ \scriptscriptstyle  {\rm I}\!{\rm N}$} }   }
\def\IK{\mathchoice{ \hbox{${\rm I}\!{\rm K}$} }
                   { \hbox{${\rm I}\!{\rm K}$} }
                   { \hbox{$ \scriptstyle  {\rm I}\!{\rm K}$} }
                   { \hbox{$ \scriptscriptstyle  {\rm I}\!{\rm K}$} }   }
\def\IQ{\mathchoice{ \hbox{${\rm l}\!\!\!{\rm Q}$} }
                   { \hbox{${\rm l}\!\!\!{\rm Q}$} }
                   { \hbox{$ \scriptstyle  {\rm l}\!\!\!\!\:{\rm Q}$} }
                   { \hbox{$ \scriptscriptstyle {\rm l}\!\!\!\!\:{\rm Q}$} }   }
\def\IC{\mathchoice{ \hbox{${\rm l}\!\!\!{\rm C}$} }
                   { \hbox{${\rm l}\!\!\!{\rm C}$} }
                   { \hbox{$ \scriptstyle  {\rm l}\!\!\!\!\:{\rm C}$} }
                   { \hbox{$ \scriptscriptstyle {\rm l}\!\!\!\!\:{\rm C}$} } }
\def\Im{\mathchoice{ \hbox{\i$\!{\rm m}$} }
                   { \hbox{\i$\!{\rm m}$} }
                   { \hbox{\i$ \scriptstyle  \!{\rm m}$} }
                   { \hbox{\i$ \scriptscriptstyle  \!{\rm m}$} }   }
\def\Ip{\mathchoice{ \hbox{${\rm p}\!\!\!\:{\rm p}$} }
                   { \hbox{${\rm p}\!\!\!\:{\rm p}$} }
                   { \hbox{$ \scriptstyle  {\rm p}\!\!\!\:{\rm p}$} }
                   { \hbox{$ \scriptscriptstyle  {\rm p}\!\!\!\:{\rm p}$} }   }


\def\IZ{\mathchoice{ \hbox{${\rm Z}\!\!\!\:{\rm Z}$} }
                   { \hbox{${\rm Z}\!\!\!\:{\rm Z}$} }
                   { \hbox{$ \scriptstyle  {\rm Z}\!\!\!\:{\rm Z}$} }
                   { \hbox{$ \scriptscriptstyle  {\rm Z}\!\!\!\:{\rm Z}$} }   }


\font\sqi=cmssqi8
\def\IRi{\hbox{$I\!\!R$}}
\def\IPi{\hbox{$I\!\!P$}}
\def\INi{\hbox{$I\!\!N$}}
\def\ICi{\hbox{$C\hskip-.7em\raise.75pt\hbox{\sqi I}\;\;$}}
\def\Imi{\hbox{$\imath\!m$}}
\def\Ipi{\hbox{$p\!\!\!\:p$}}

\def\Rmop#1{ {\rm #1}}


\def\kasten#1{\mathop{\mkern0.5\thinmuskip
                      \vbox{\hrule
                            \hbox{\vrule
                                  \hskip#1
                                  \vrule height#1 width 0pt
                                  \vrule}%
                            \hrule}%
                      \mkern0.5\thinmuskip}}

\def\Kasten{\mathchoice{\kasten{8pt}}%
                       {\kasten{6pt}}%
                       {\kasten{4pt}}%
                       {\kasten{3pt}}}%

\begin{center}
{\Large \bf On the Flux-Across-Surfaces Theorem}\\
\bigskip

M.~Daumer, D.~D\"urr\\
{\it Fakult\"at f\"ur Mathematik, Universit\"at M\"unchen,
      Theresienstr. 39, 80333 M\"unchen, Germany}\\
S.~Goldstein\\
{\it Department of Mathematics, Rutgers University, New Brunswick,
      New Jersey 08903, USA}\\
 N.~Zangh\`{\i}\\ {\it Istituto di Fisica dell'Universit\`a di Genova, INFN, Via
Dodecaneso 33, 16146 Genova, Italy} \end{center}
 
\date{\today}

 
\section*{Abstract}  The quantum probability flux of a particle integrated
over time and a distant surface gives the probability for the particle 
crossing that surface at some time.
 We prove  the free \FAS , which was conjectured by Combes,  Newton and
Shtokhamer \cite{Combes}, and which relates the integrated  quantum flux to
the usual quantum mechanical formula for the cross section.  The integrated
quantum flux is equal to the probability of outward crossings of 
surfaces by Bohmian trajectories in the scattering regime.

\section{Introduction}

Time-dependent scattering theory is concerned  with the long-time behavior
of wave packets $\psi_t$.
Dollard's \SIC\ \cite{Dollard,Dollard2} 
asserts that, assuming, say, asymptotic completeness, the probability of finding a particle with a wave function
$\psi\in\eitsch_{ac}(H)$,  the absolutely continuous subspace for the
Hamiltonian $H$, 
in the far future in a
given cone $C\subset \IR^3$ (with vertex at the origin) equals the
probability that the quantum mechanical momentum of $\Omega_-^\dagger\psi$
lies in  the same cone, \be \label{SICT} \lit \int_C d^3x |\psi_t(\bx)|^2 =
\int_C d^3v |\widehat{\Omega_-^\dagger\psi}(\bv)|^2, \ee where $\Omega_-:=
\slim_{t\to\infty} e^{iHt}e^{-iH_0t}$ is the   wave operator, $H = H_0 +V$
with the free Hamiltonian $H_0 = -\Delta/2$  (we choose units such that $\hbar=m=1$) and the interaction potential $V$. $\widehat{\ }$
denotes the Fourier transform.  The \SIC\ is regarded as fundamental,  from
which  the expression for the differential
 cross section $\frac{d\sigma}{d\Omega}=
|f(\theta,\phi)|^2$ from the time-independent theory is to be derived 
from the r.h.s.\ of (\ref{SICT})
(e.g.\ \cite{ReedSimon3}, p.\ 356, 
\cite{Enss2}).

Combes, Newton and Shtokhamer \cite{Combes} observed however 
that what is relevant for  scattering theory  is 
 a formula for the probability that the particle  crosses some distant
surface  at some time during the scattering process, since the detectors
click at some {\it random} time, which is {\it not} chosen by the
experimenter. Heuristically, this probability should be given by integrating
the quantum mechanical probability  flux over the relevant time interval and
this surface. (The flux is often used that way in  textbooks.)
 Combes, Newton and  Shtokhamer hence conjectured the ``\FAS''  \be
\label{FAST} \lir \int_0^\infty dt \int_{C\cap\PKR} \bj^{\psi_t} \cdot \bn
d\sigma= \int_C d^3v |\widehat{\Omega_-^\dagger\psi}(\bv)|^2, \ee where
$B_R$ is the ball with radius $R$ and outward normal $\bn$. To our knowledge
there exists no proof of this theorem. A simpler statement, also not
previously proven, is the ``free \FAS ,'' for freely evolving $\psi_t$, \be
\label{FASFREET} \lir \int_0^\infty dt \int_{C\cap\PKR} \bj^{\psi_t} \cdot
\bn d\sigma= \int_C d^3v |\psihat(\bv)|^2 \ee which in a sense is physically
good enough, because the scattered wave packet will  move almost freely  
after the scattering has essentially been completed (see also
\cite{Combes}).  We shall prove the ``free \FAS '' in this paper, commenting
at the end on the  general \FAS .

We want first to give the heuristic argument for (\ref{FASFREET}). The flux
should contribute to the integral in (\ref{FASFREET}) only for large times,
because the packet has to travel a long time before it reaches the  distant
sphere $\PKR$, so that we may use the long-time asymptotics  of the free
evolution. Writing \be \label{FREEHEU} \psi_t(\bx) = (e^{-iH_0t} \psi)(\bx)
= \int d^3y \frac{ e^{i\frac{|\bx-\by|^2}{2t}} }{(2\pi it)^{3/2}} \psi(\by)
\ee and expanding the exponent of the propagator, we obtain \be \psi_t(\bx)
= \frac{e^{i\xxt}}{(it)^{3/2}} \hat{\psi}(\xt) +
\frac{e^{i\xxt}}{(it)^{3/2}} \int \frac{d^3y}{(2\pi)^{3/2}}e^{-i\xyt} 
(e^{i\yyt} - 1) \psi(\by) \ee so that for large times (the second term
should be negligible since $|(e^{i\yyt} - 1)| \to  0$ as $t\to\infty$) \be
\label{ASYMP} \psi_t(\bx) \approx (it)^{-3/2}e^{i\xxt} \hat{\psi}(\xt). \ee
The importance of this asymptotics  for scattering theory has long been
recognized, see e.g.\ \cite{Brenig} and \cite{Dollard}.

Consider now a cone $C$. Substituting $\bv := \frac{\bx}{t}$ one readily
obtains the  \SIC\  \be \label{SIC} \lit \int_{C} d^3x |\psi_t(\bx)|^2 =
\int_{C} d^3v |\psihat(\bv)|^2. \ee But the l.h.s.\ of (7) should be
unaffected if $C$ is replaced by the truncated cone $C_R=C\cap B_R^c,
B_R^c := \IR^3 \setminus B_R$, for
any $R>0$. Thus writing $\int_{C_R} d^3x  |\psi_t(\bx)|^2 = \int_0^t dt'
\int_{C_R} d^3x \frac{\partial}{\partial t'} |\psi_{t'}(\bx)|^2 + \int_{C_R}
d^3x  |\psi_0(\bx)|^2$ and using the quantum flux equation
$\frac{\partial}{\partial t} |\psi_t|^2 + \nabla \cdot \jpsi = 0$ together
with
 Gauss' theorem and taking $R\to\infty$ provides a heuristic argument for
the free \FAS . Unfortunately, because of the difficulty in controlling the
relevant approximations, this  argument cannot be readily turned into a
rigorous proof (see also \cite{Combes}).

Instead we may more directly compute the flux using
(\ref{ASYMP}), from which  we find for  $t\to\infty$
\be
\label{JASYMPHEU}
\jpsi(\bx) = {\mbox{\rm Im}}  \/ \psi_t^*(\bx) \nabla \psi_t(\bx) \approx
 t^{-3} |\hat{\psi}(\xt)|^2\xt.  \ee Noting  that the flux is purely
outgoing for large times, i.e.\ parallel to the outward normal $\bn$ of
$\PKR$, we then find upon substituting $\bv := \xt$ that   \ba \int_0^\infty
dt \int_{\CKR} \jpsi \cdot \bn d\sigma  &\approx&  \int_0^\infty dt
\int_{\CKR} t^{-3}|\psihat(\xt)|^2   \xt \cdot \bn(\bx) d\sigma \nonumber \\
&=& \int_C d^3v |\psihat(\bv)|^2. \ea 
(Note that the middle term above does
not in fact depend upon $R$.)
This calculation  can smoothly be
turned into a rigorous proof, to which we now turn.

\newpage

\section{The Flux-Across-Surfaces Theorem}

First we fix the following  notation,
illustrated also in the figure.

\begin{figure}[h]
\label{STREUNG}
\begin{center}
\leavevmode%
\epsfxsize=10cm
\epsffile{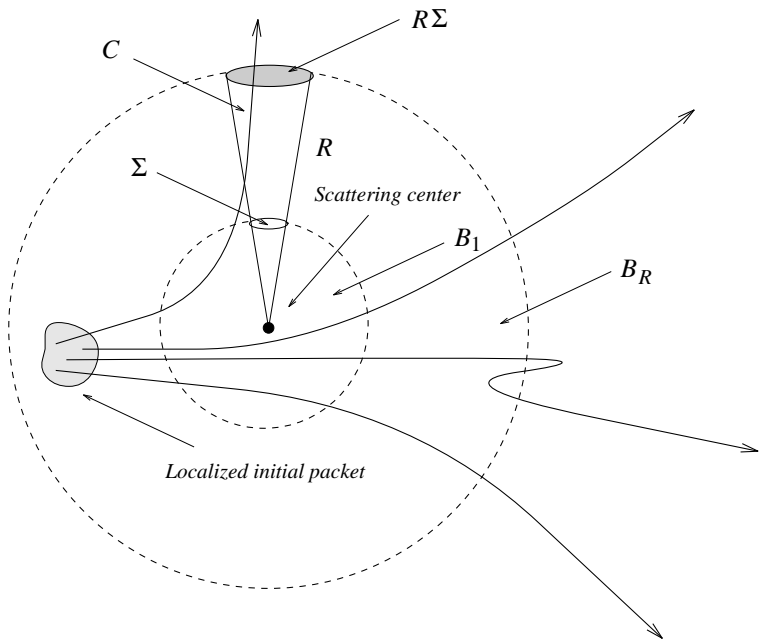}
\end{center}
\caption{The initial wave packet evolves under the influence of the
scatterer at the origin. In Bohmian mechanics (see remark 3.11) 
the flow lines of the corresponding flux 
represent the possible trajectories of the particle.}
 \end{figure}

For $R>0$ let
$B_R := \{\bx\in\IR^3: x \le R \}$ and $\PKR = \{\bx\in\IR^3: x = R \}$,
with $x=|\bx|$.
 Further let $\bn : \PKR \to \IR^3$,
$\bn(\bx) := \frac{\bx}{x}$  be the outward normal of the sphere
  $\PKR$. The cone spanned by the subset
 $\Sigma \subset \partial B_1$ of the unit sphere    is
$C := \{\lambda \bx \in \IR^3: \bx \in \Sigma, \lambda \ge 0 \}$ and its
intersection with the sphere
  $\PKR$
is $R\Sigma := C \cap \PKR = \{R\bx \in \IR^3: \bx \in \Sigma \}$. 
Another characterization of cones is provided by
the unit vector $\bn_C$, $\|\bn_C\|=1$ and the opening angle
  $\theta_C \in [0,\pi]$, namely  $C:= \{\bx \in \IR^3:
\bx \cdot \bn_C > x \cos \theta_C \}$. 
We chose polar coordinates
 $(r,\theta, \phi), r\ge 0, \theta \in [0,\pi], \phi \in [0,2\pi)$ centered
at the origin,  $\bx(r,\theta,\phi) = (r\sin \theta \cos \phi,
r\sin\theta\sin\phi, r\cos\theta),$ with the $z$-direction $\bn_C$. In these
polar coordinates $B_R = \{(r, \theta, \phi): r \le R \}$, $\PKR = \{(r,
\theta, \phi): r=R \}$ and  $C = \{(r,\theta, \phi): \theta < \theta_C \}$. 
The intersection of the cone  $C$ with the sphere  $\PKR$  is now $C\cap
\PKR = \{(r,\theta,\phi): r=R, \theta < \theta_C \}$ with outward normal
 $\bn(\theta,\phi) = R^{-1} \bx(R,\theta,\phi)$. $d\Omega =
\sin\theta d\theta d\phi$ denotes the solid angle.

\begin{theorem}
\label{FASFREE}

Let $\psi \in {\cal S}(\IR^3)$ and $\psi_t := e^{-iH_0t}\psi$. 
Then for all $T\in \IR$ and any cone ~$C$ 
\be
\label{FASFREEF}
\lir \int_T^\infty dt \int_\CKR \jpsi(\bx) \cdot \bn d\sigma
= \lir \int_T^\infty dt \int_\CKR |\jpsi(\bx) \cdot \bn |d\sigma
=\int_C d^3v |\psihat(\bv)|^2.\nonumber
\ee
\end{theorem}

\begin{rmrem}
The condition
  $\psi\in {\cal S}(\IR^3)$, the Schwarz space,
 is introduced for the sake of simplicity. The  proof may be 
performed with milder assumptions. Note, however, that  ${\cal S}(\IR^3)$
is a time invariant domain under the free evolution.
\end{rmrem}
\begin{rmrem}
The reason for formulating the theorem as slightly stronger than (\ref{FASFREET}),
including information also about the modulus of $\jpsi \cdot \bn$,
is that  in Bohmian mechanics (see remark \ref{BOHMREM}) the first (second) 
flux integral in (\ref{FASFREEF})
gives simply the expected value of the number of signed crossings
(the total number of crossings) by the
Bohmian trajectories of the surface.
If they both agree it is an easy consequence that
(\ref{FASFREEF}) equals the  asymptotic probability that the particle crosses
$C\cap \PKR$ at some time in $[0,\infty)$. 
\end{rmrem}

It will be convenient to introduce a notion of closeness of 
fluxes.

\begin{rmdef}
\label{FASCLOSE}
Two smooth functions $\bj_1,\bj_2 : \IR^3\times\IR \to \IR^3$ are said to
be  ``close in the sense of the asymptotic flux across surfaces,'' or
$\bj_1 \stackrel{FAS}{\sim} \bj_2$, if for some $T>0$
\be
\label{CONDFASCLOSE}
\lir \int_T^\infty dt \int_\PKR |(\bj_1-\bj_2)\cdot \bn|d\sigma = 0.
\ee
\end{rmdef}

\begin{lemma} \label{CLASSICFLUX} Suppose that  for $\bj: \IR^3\times\IR \to
\IR^3$ and $\bj^\Phi_0(\bx,t) :=  t^{-3} |\Phi(\xt)|^2\xt$ with smooth $\Phi
\in L^2(\IR^3)$, we have $\bj \stackrel{FAS}{\sim} \bj^\Phi_0$. Then for all
cones  $C\subset \IR^3$ and some $T>0$  \be \label{ALL} \lir \int_T^\infty
dt \int_\CKR \bj (\bx,t) \cdot \bn(\bx) d\sigma = \lir \int_T^\infty
dt \int_\CKR |\bj(\bx,t) \cdot \bn(\bx)| d\sigma =  \int_C  d^3v
|\Phi(\bv)|^2. \ee \end{lemma}

{\bf Proof:} 
By  definition (\ref{FASCLOSE}) it is sufficient to establish   (\ref{ALL}) for
$\bj$ replaced by $\bj^\Phi_0$.

Using spherical coordinates  $\bx(r,\theta,\phi) =  (r\sin \theta \cos \phi,
r\sin\theta\sin\phi, r\cos\theta)$ we compute \ba \int_T^\infty dt \int_\CKR
\bj^\Phi_0 (\bx) \cdot \bn(\bx) d\sigma &=& \int_T^\infty dt\int_\CKR t^{-3}
 |\Phi(\xt)|^2 \xt \cdot \bn(\bx) d\sigma \nonumber \\ &=& \int_T^\infty
dt\int_\Sigma d\Omega R^2 t^{-3}  |\Phi(\frac{\bx(R,\theta,\phi)}{t})|^2
\frac{\bx(R,\theta,\phi)}{t} \cdot \bn(\theta,\phi). \nonumber \ea Observing that
 $\frac{\bx(R,\theta,\phi)}{t} = \bx(\frac{R}{t},\theta,\phi)$ and
substituting  $v := \frac{R}{t}$ we obtain \ba \lir \int_T^\infty dt
\int_\CKR \bj^\Phi_0 (\bx) \cdot \bn(\bx) d\sigma &=& \lir \int_0^{R/T} dv
v^2 \int_\Sigma d\Omega |\Phi(v,\theta,\phi)|^2 \nonumber \\ &=& \int_C
d^3v |\Phi(\bv)|^2.   \ea The observation that $\bx \cdot \bn(\bx) =  |\bx
\cdot \bn(\bx)|$ finally shows that all equalities in (\ref{ALL}) hold.

\proofend

\begin{lemma} \label{FREELEMMA} Let $\psi \in {\cal S}(\IR^3), \psi_t :=
e^{-iH_0t}\psi$ and $\jpsi = {\rm Im} \psi_t^* \nabla \psi_t$. Then  \be
\label{CONDFREELEMMA} \jpsi(\bx,t) \stackrel{FAS}{\sim} \xt t^{-3}
|\psihat(\xt)|^2. \ee

\end{lemma}

{\bf Proof:}  We verify the conditions in definition (\ref{FASCLOSE}). For
$t>0$ we may write \ba \label{FREEASYMPTOTICS} \psi_t(\bx) &=& (e^{-iH_0t}
\psi)(\bx) \nonumber \\ &=&\int d^3y \frac{ e^{i\frac{|\bx-\by|^2}{2t}}
}{(2\pi it)^{3/2}} \psi(\by)
 \nonumber \\ &=& \frac{e^{i\xxt}}{(it)^{3/2}} \hat{\psi}(\xt) +
\frac{e^{i\xxt}}{(it)^{3/2}} \int \frac{d^3y}{(2\pi)^{3/2}}e^{-i\xyt}
(e^{i\yyt} - 1) \psi(\by). \ea Since \be \label{EKLEIN}
 |e^{i\yyt} - 1| \le 2 \ee for all $\by\in\IR^3, t>0$, we obtain that \be
\label{FDEF} f(\bv,t) := \int \frac{d^3y}{(2\pi)^{3/2}} e^{-i\bv\cdot\by}
(e^{i\yyt} - 1)\psi(\by)   \ee is well defined for all $\bv\in\IR^3$.
Because  $\psi \in {\cal S}(\IR^3)$  we may interchange differentiation and
integration to further  obtain that $f$ is differentiable on $\IR^3 \times
[T,\infty)$. 

It is useful to introduce \be \label{GDEF} \bg(\bv,t) := \nabla f(\bv,t) =
-i \int \frac{d^3y}{(2\pi)^{3/2}} e^{-i\bv\cdot\by} (e^{i\yyt} - 1) \by
\psi(\by).  \ee Further we put  \be \alpha(\bx,t) :=
\frac{e^{i\xxt}}{(it)^{3/2}} \hat{\psi}(\xt) \ee and \be \beta(\bx,t) :=
\frac{e^{i\xxt}}{(it)^{3/2}} f(\xt,t), \ee i.e.\ $\psi_t(\bx) =
\alpha(\bx,t) + \beta(\bx,t)$, and \ba \nabla \alpha(\bx,t) &=&
\frac{e^{i\xxt}}{(it)^{3/2}} \left( i\xt \psihat(\xt) + \frac{1}{t} (\nabla
\psihat)(\xt) \right) \\ \nabla \beta(\bx,t) &=&
\frac{e^{i\xxt}}{(it)^{3/2}} \left( i\xt f(\xt,t) +  \frac{1}{t} \bg(\xt,t)
\right). \ea We may thus write \ba \label{CURRENT} \jpsi(\bx) &=& {\mbox{\rm
Im}} (\psi_t^*(\bx) \nabla \psi_t(\bx) ) \nonumber \\ &=& {\mbox{\rm Im}}
(\alpha^*(\bx,t) \nabla \alpha(\bx,t) + \beta^*(\bx,t) \nabla \alpha(\bx,t)+
\alpha^*(\bx,t)  \nabla \beta(\bx,t) + \beta^*(\bx,t) \nabla \beta(\bx,t))
\nonumber \\ &=& \frac{\bx}{t} t^{-3} |\psihat(\frac{\bx}{t})|^2 + N(\bx,t),
\ea with \be \label{NDEF} N(\bx,t) := {\mbox{\rm Im}} \Bigl( t^{-4}
\psihat^*(\xt)\nabla\psihat(\xt) + \beta^*(\bx,t) \nabla \alpha(\bx,t) +
\alpha^*(\bx,t) \nabla \beta(\bx,t) + \beta^*(\bx,t) \nabla
\beta(\bx,t)\Bigr). \ee

Thus to obtain (\ref{CONDFREELEMMA})  we need only  show that
(\ref{CONDFASCLOSE}) is satisfied for some  $T>0$ 
and $\bj_1-\bj_2$ given by (\ref{NDEF}).
We shall make use of the bounds
\ba
\label{FGLEMMA}
\sup_{\bv\in\IR^3, t>0}  |f(\bv,t)|  \le 2(2\pi)^{-3/2} \|\psi\|_1 =: c_f, \\
\label{GLEMMA}
\sup_{\bv\in\IR^3, t>0}  |g(\bv,t)|  \le 2(2\pi)^{-3/2}\|y\psi(\by)\|_1 =: c_g
\ea
($\| \cdot \|_1$ denotes the norm in $L_1$)
and the fact  that
\be
\label{FTOZERO}
\lir f(\bv,\frac{R}{v}) = 0 \quad \forall \bv\in\IR^3.
\ee
(Note that $f(\bv,\frac{R}{v})$ is well defined even for $\bv=0$ by
(\ref{FDEF}).)
(\ref{FGLEMMA}) and (\ref{GLEMMA}) 
hold since  $|e^\frac{iy^2}{2t} - 1| \le 2$ for all $\bv,\by\in\IR^3, t>0$. 
Since
$\psi \in L_1(\IR^3)$ and $\lir |e^{i\frac{y^2v}{2R}} -1| = 0$ for all 
$\bv,\by \in \IR^3$, (\ref{FTOZERO}) follows by dominated convergence. 

  We analyze the contribution
of the expressions on the r.h.s.\ of
(\ref{NDEF}) term by term. For the first term we obtain, using 
 $|{\mbox{\rm Im}} z| \le |z|$, the substitution  $v = \frac{R}{t}$, and the
Schwarz inequality \ba \label{TERM1} |\IM \int_T^\infty dt \int_\PKR t^{-4}
\psihat^*(\xt) (\nabla \psihat)(\xt) \cdot \bn d\bsigma | &\le& \int_T^\infty dt
\int_{\partial B_1} d\Omega R^2 t^{-4}  |\psihat(\xt)||(\nabla \psihat)(\xt)| 
\nonumber \\&\le& \int_0^\infty dv v^2\int_{\partial B_1} d\Omega R^{-1} 
|\psihat(\bv)||(\nabla \psihat)(\bv)| \nonumber \\ &\le& R^{-1} \|\psihat\|_2
\||\nabla \psihat| \|_2  \to 0 \ea as $R\to\infty$, since $\psihat \in {\cal
S}$.

For the second term 
\ba
\label{TERM2A}
\IM \beta^*\nabla\alpha &=& \IM t^{-3} f^*(\xt,t) \left(i\xt \psihat(\xt) +
\frac{1}{t} (\nabla \psihat)(\xt) \right)
\ea
we obtain, similarly using
 (\ref{FGLEMMA}), \ba \label{TERM2} |\IM \int_T^\infty dt \int_\PKR \beta^*
\nabla \alpha \cdot \bn  d\sigma|  &\le& \int_T^\infty dt\int_{\partial B_1}
d\Omega R^2t^{-3} |f^*(\xt,t)|\bigl(|\frac{R}{t}\psihat(\xt)| + \frac{1}{t}
|(\nabla \psihat)(\xt) | \bigr)\nonumber \\ &\le& \int_0^\infty dv
\int_{\partial B_1} d\Omega v^2 |f^*(\bv,\frac{R}{v})|\bigl( |\psihat(\bv)| +
\frac{1}{R} |(\nabla \psihat)(\bv) |\bigr) \nonumber \\ &\le& \int_0^\infty dv
\int_{\partial B_1} d\Omega v^2 |f^*(\bv,\frac{R}{v})| |\psihat(\bv)| +
\frac{1}{R} c_f \||\nabla \psihat| \|_1. \nonumber \ea The second term tends
to zero as   $R\to\infty$, and the first term  also vanishes: using
(\ref{FGLEMMA})
 and the fact that $\psihat\in L^1(\IR^3)$ we see that the integrand is
dominated by an integrable function uniformly in $R$, so that with 
(\ref{FTOZERO}) the integral vanishes for $R\to\infty$ by dominated
convergence.
 
For  \ba \IM \alpha^*\nabla\beta &=& \IM t^{-3} \psihat^*(\xt) \left( i\xt
f(\xt,t) + \frac{1}{t} g(\xt,t) \right)  \ea we may proceed in an analogous
manner and obtain \ba \label{TERM3} |\IM \int_T^\infty dt\int_\PKR
\alpha^*\nabla\beta \cdot \bn d\sigma|  &\le& \int_T^\infty dt\int_{\partial B_1}
d\Omega R^2t^{-3} |\psihat^*(\xt)| (|\frac{R}{t} f(\xt,t)| +
\frac{1}{t}|\bg(\xt,t)|) \nonumber \\ &\le& \int_0^\infty dv \int_{\partial B_1}
d\Omega v^2 |\psihat^*(\bv)| (|f(\bv,\frac{R}{v})| +
\frac{1}{R}|\bg(\bv,\frac{R}{v})|) \nonumber \\ &\le& \int_0^\infty dv
\int_{\partial B_1} d\Omega v^2 |\psihat^*(\bv)| |f(\bv,\frac{R}{v})| + c_g
\frac{1}{R}\| \psihat \|_1 \nonumber \\ &\to& 0 {{\rm ~ as} } \quad R \to
\infty. \ea

It remains to show that for some $T>0$
\be
\lir \int_T^\infty dt \int_\PKR |\beta^* \nabla \beta \cdot \bn | d\sigma = 0.
\ee
Now,
\ba
\beta^*\nabla \beta &=& it^{-3} f^*(\xt,t) \int \frac{d^3y}{(2\pi)^{3/2}}
e^{-i\xt \cdot \by} (e^{i\yyt} -1 )(\xt - \yt) \psi(\by)    \\
&=&   -it^{-3} f^*(\xt,t)  \int \frac{d^3y}{(2\pi)^{3/2}}
e^{-i\xt \cdot \by}(\xt - \yt) \psi(\by)  \nonumber \\
&+& it^{-3} f^*(\xt,t)  \int \frac{d^3y}{(2\pi)^{3/2}}
e^{-i\xt \cdot \by} e^{i\yyt} (\xt - \yt) \psi(\by) \nonumber  \\
\label{VORVORLETZTER}
&=& t^{-3} f^*(\xt,t)  \int \frac{d^3y}{(2\pi)^{3/2}} \Bigl( \nabla_{\by}
e^{-i\frac{\bx}{t}\cdot \by} \Bigr) \psi(\by)   \\ \label{ZWISCHEN} 
&-& t^{-4} f^*(\xt,t) (\nabla \psihat)(\xt)    \\
\label{LETZTER}
&-& t^{-3}  f^*(\xt,t)  
\int \frac{d^3y}{(2\pi)^{3/2}}
\nabla_{\by} \Bigl( e^{-i\xt \cdot \by} e^{i\yyt} \Bigr) 
\psi(\by).
\ea
Treating (\ref{ZWISCHEN}) like   (\ref{TERM2A}) we see that (\ref{ZWISCHEN})
doesn't contribute. Partial  integration of  (\ref{VORVORLETZTER}) $+$ 
(\ref{LETZTER}) 
yields
\ba 
{\bf a}(\bx,t) &:=& t^{-3}  f^*(\xt,t)  
\int \frac{d^3y}{(2\pi)^{3/2}}
\Bigl( e^{-i\xt \cdot \by} e^{i\yyt}  
\nabla_{\by} \psi(\by) -  e^{-i\xt \cdot \by} \nabla_{\by} \psi(\by) \Bigr) 
\nonumber \\
&=& t^{-3} f^*(\xt,t) \int  \frac{d^3y}{(2\pi)^{3/2}}
e^{-i\xt\cdot\by} (e^{i\yyt} - 1) \nabla_{\by} \psi(\by) \nonumber \\
&=& -t^{-3} f^*(\xt,t) \int \frac{d^3y}{(2\pi)^{3/2}} 
\frac{t^2}{x^2} (\nabla_{\by}^2 e^{-i\xt\cdot\by})
(e^{i\yyt} - 1) \nabla_{\by} \psi(\by) \nonumber \\
&=& -t^{-1} x^{-2} f^*(\xt,t) \int \frac{d^3y}{(2\pi)^{3/2}} 
  e^{-i\xt\cdot\by} \nabla_{\by}^2\Bigl(
(e^{i\yyt} - 1) \nabla_{\by} \psi(\by) \Bigr),
\ea 
with  two partial integrations in the last step.

Now
\ba
\nabla_{\by} (e^{i\yyt} - 1) &=& it^{-1}\by e^{i\yyt}, \nonumber \\ 
\nabla_{\by}^2(e^{i\yyt} - 1)  &=& (-\frac{y^2}{t^2} + 3it^{-1})e^{i\yyt}
\ea
and
\be
|e^{i\yyt} -1| \le \yyt,
\ee
so that for $t\ge T > 0$   
\be
| \nabla_{\by}^2(
(e^{i\yyt} - 1) \nabla_{\by} \psi(\by) )| \le |h(\by)| t^{-1}
\ee
with some $h\in {\cal S}(\IR^3)$ appropriately chosen.
Hence,
\be
|\int \frac{d^3y}{(2\pi)^{3/2}} 
  e^{-i\xt\cdot\by} \Bigl(\nabla_{\by}^2( (e^{i\yyt} - 1) \nabla_{\by}
\psi(\by) )\Bigr)|  \le  t^{-1} \int \frac{d^3y}{(2\pi)^{3/2}} |h(\by)| =: 
c t^{-1}. \ee Thus we arrive at  \be |{\bf a}(\bx,t)| \le  c t^{-2} x^{-2}
|f(\xt,t)|, \ee and with $\br(\theta,\phi) := \bx(R,\theta,\phi)$ we obtain
\ba \label{NAJA} \int_T^\infty dt \int_\PKR |{\bf a} \cdot  \bn|d\sigma
&\le&
 c \int_T^\infty dt t^{-2} \int_{\partial B_1} d\Omega  |f(\frac{\br}{t},t)|. \ea 
On the one hand  (cf.\ (\ref{FGLEMMA})) \be \sup_{t\ge T, R>0}
|f(\frac{\br}{t},t)| \le c_f, \ee and on the other hand with the 
Riemann-Lebesgue lemma  \ba  \lir  |f(\frac{\br}{t},t)| = 0 \ \forall \ t>0.
\ea Hence the r.h.s. of (\ref{NAJA}) tends to zero (dominated convergence)
as $R\to\infty$ and we have thus finished the  the proof of lemma
(\ref{FREELEMMA}). \proofend

\begin{corollary}
\label{FASTOK}
For some $T>0$,  Theorem (\ref{FASFREE}) holds . 
\end{corollary}

The analysis so far actually establishes the theorem for any $T>0$. We now
show that the restriction $T>0$ can be removed.

\begin{lemma}
\label{FINITELEMMA}
For all $-\infty < T_1 < T_2 < \infty$ 
\be
\lir \int_{T_1}^{T_2} dt \int_{\PKR} |\jpsi(\bx)\cdot\bn|d\sigma = 0.
\ee
\end{lemma} 

{\bf Proof}:  First observe that \be \label{DOLL} \int_{T_1}^{T_2} dt
\int_{\PKR} |\jpsi(\bx)\cdot\bn|d\sigma \le 4\pi  \int_{T_1}^{T_2} dt R^2
\sup_{\bx\in \PKR} |\psi_t(\bx)||\nabla\psi_t(\bx)|. \ee We want to apply
dominated convergence. With   \be \label{PAIN} \psi_t(\bx) = (2\pi)^{-3/2}
\int d^3k
 e^{i\bk\cdot\bx}e^{-i\kkt} \psihat(\bk),
\ee
and 
\be
\nabla \psi_t(\bx) = i(2\pi)^{-3/2} \int d^3k
 e^{i\bk\cdot\bx}e^{-i\kkt} \bk \psihat(\bk),
\ee
 we have that \be \sup_{\bx\in\IR^3, t\in\IR} |\nabla \psi_t(\bx)| \le
(2\pi)^{-3/2} \|k\psihat(\bk)\|_1. \ee Since  $\psi\in {\cal S}(\IR^3)$ we
may perform $n$ partial integrations in (\ref{PAIN}) to obtain    \ba
\psi_t(\bx) &=&  (2\pi)^{-3/2} x^{-n} \int d^3k
\Bigl[(\frac{1}{i}\nabla_k)^n e^{i\bk\cdot\bx}\Bigr] e^{-i\kkt} \psihat(\bk)
\nonumber \\ &=& (-1)^n (2\pi)^{-3/2} x^{-n}  \int d^3k  e^{i\bk\cdot\bx}
\Bigl[(\frac{1}{i}\nabla_k)^n e^{-i\kkt} \psihat(\bk)\Bigr]. \ea We 
estimate  \be |(\frac{1}{i}\nabla_k)^n e^{-i\kkt} \psihat(\bk)| \le |h(\bk)|
(1+t^n) \ee for some $h \in {\cal S}(\IR^3)$. For $n=2$ we thus have \ba
 R^2 \sup_{\bx\in \PKR} |\psi_t(\bx)||\nabla\psi_t(\bx)| &\le& (2\pi)^{-3}
\|k\psihat(\bk)\|_1 \int d^3k |(\frac{1}{i}\nabla_k)^2  e^{-i\kkt}
\psihat(\bk)| \nonumber \\ &\le& c' (1+t^2)  \in L^1(T_1,T_2). \ea For $n=3$
and any fixed $t\in[T_1,T_2]$ we obtain \ba R^2 \sup_{\bx\in \PKR}
|\psi_t(\bx)||\nabla\psi_t(\bx)| &\le& (2\pi)^{-3}
R^{-1}\|k\psihat(\bk)\|_1 \int d^3k |(\frac{1}{i}\nabla_k)^3  e^{-i\kkt}
\psihat(\bk)| \nonumber \\ &\le& c'' R^{-1} (1+t^3) \to 0 \ea for
$R\to\infty$.  Now we use dominated convergence in (\ref{DOLL}) and are
done.  \proofend

Theorem (\ref{FASFREE}) now follows directly from Cor.(3.7) and Lemma
(\ref{FINITELEMMA}).

\begin{rmrem}

The extension of our result to the free evolution of $N$ particles is
straightforward.  The extension to the interacting case, i.e.\ a proof of
(\ref{FAST}) (even for one-particle scattering), is open. The theory of
generalized eigenfunction expansions \cite{Ikebe} can be used to control the
 space-time behavior of $\psi_t(\bx)$ and of the flux $\jpsi$. We may expand
$\psi_t(\bx) =  (2\pi)^{-3/2} \int d^3k e^{-i\kkt} \phi(\bx,\bk)
\widehat{\Omega_-^\dagger\psi}(\bk)$,  where $\phi(\bx,\bk)$ are solutions
of the Lippmann-Schwinger equation  \be \label{LS1} \phi(\bx,\bk) =
e^{i\bk\cdot\bx} -  \frac{1}{2\pi} \int d^3y
\frac{e^{-i{k|\bx-\by|}}}{|\bx-\by|} V(\by) \phi(\by,\bk), \ee (with {\it
incoming} spherical waves). The important connection between the wave
operators, generalized eigenfunctions and the Fourier transform is 
expressed by
 $\widehat{\Omega_-^\dagger\psi}(\bk) = (2\pi)^{-3/2}\int d^3x
\phi^*(\bx,\bk) \psi(\bx)$.  For a proof of (\ref{FAST}), relying
essentially on a stationary phase argument, we need additional smoothness
properties  of the eigenfunctions  which, to our knowledge, have not yet
been established. More precisely, we need to know that 
$\phi(\bx, \cdot) \in C^\infty(\IR^3\setminus \{0\} )$ for all $\bx\in\IR^3$,
 $\phi(\cdot,\bk) \in C^\infty(\IR^3) $ for all $\bk\in\IR^3\setminus \{0\}$,
and $\sup_{\bx\in \IR^3, \bk \in
\IR^3\setminus \{0\} } \phi(\bx,\bk) < \infty$. The closest we could get
was, with \cite{Ikebe} and \cite{ReedSimon3}  Theorem XI.41 and XI.70,  that
for
 $V \in L^2(\IR^3)$ locally H\"older continuous with the possible exception
of finitely many  singularities and $|V(\bx)| = O(x^{-2-h})$ for some $h>0$, 
 $\phi(\bx,\bk)$ is bounded and continuous for $\bx \in \IR^3$ and $\bk \in D
\subset \IR^3\setminus \{0\}$, where $D$ is {\it compact}.  It is well known
that for
 $V\in C^\infty(\IR^3)$  the solutions $\phi$ of the 
stationary \se\
obey $\phi \in C^\infty(\IR^3)$ and thus the solutions $\phi(\cdot, \bk)$
of the Lippmann-Schwinger equation, which are special solutions of the 
stationary \se\
parametrized by $\bk$, are in $C^\infty(\IR^3)$
(see \cite{ReedSimon2}, Theorem IX.62). It remains to be shown that  for any
$\bx\in\IR^3$ both $\phi(\bx,\cdot) \in C^\infty(\IR^3\setminus \{0\})$ and
$\sup_{\bx\in\IR^3,\bk\in\IR^3\setminus \{0\}} |\phi(\bx,\bk)| < \infty$. 
This should be true for potentials which are sufficiently smooth and
have sufficiently strong decay at infinity \cite{DaumerPhD}. \end{rmrem}

\begin{rmrem} The mathematical physics of  scattering theory is mainly
concerned with the existence and asymptotic completeness of wave operators
$\Omega_\pm := \slim_{t\to\mp \infty} e^{iHt}e^{-iH_0t}$. The wave operators
may be used to control the long-time behavior of wave packets $\psi_t :=
e^{-iHt}\psi$, in the sense of $\psi_t \stackrel{L^2}{\sim} e^{-iH_0t}
\Omega_-^\dagger \psi$,  i.e.\ the difference vanishes in $L^2$ as
$t\to\infty$.  Dollard's lemma implies  that for $\phi_t := e^{-iH_0t}\phi$ 
\be \phi_t(\bx) \stackrel{L^2}{\sim} e^{i\xxt}(it)^{-3/2} \phihat(\xt). \ee
Asymptotic completeness of the wave operators implies, among other things,
that for any $\psi\in \eitsch_{ac}(H)$ there is a $\phi \in L^2$
such that $\lit \|e^{-iHt}\psi - e^{-iH_0t}\phi\|_2 = 0$, where
$\phi = \Omega_-^\dagger \psi$ with  $\Omega_-$  unitary on 
$\eitsch_{ac}(H)$
(see, e.g.,\ \cite{ReedSimon3}). 
It then follows
 by the
triangle inequality that for any $\psi\in\eitsch_{ac}$
\be \psi_t(\bx) \stackrel{L^2}{\sim}
e^{i\xxt}(it)^{-3/2} \widehat{\Omega_-^\dagger\psi}(\xt). \ee  From this the
general \SIC\ (\ref{SICT}) follows easily (see. e.g. \cite{Dollard}).  This
is however not sufficient to prove the physically relevant \FAS . The 
notion of closeness which should be used here is the closeness of fluxes in
the sense of the asymptotic flux across surfaces introduced in definition
(\ref{FASCLOSE}), and not the closeness of wave functions in $L^2$. 

\end{rmrem}

\begin{rmrem} \label{BOHMREM} In the context of Bohmian mechanics
\cite{Albert,Bell,BOHI,DGZ,Holland}, a theory of point particles moving
along trajectories defined by an ODE arising from the wave function, with
velocity $\jpsi/|\psi_t|^2$, a theory that  can be shown to underly the
quantum  formalism (see. e.g.\ \cite{Bohm,OPAP}), it follows easily from 
Theorem (\ref{FASFREE}) that  \be \label{IMPO} \lir \IP^\psi(\bx_e^R \in
R\Sigma) = \lir \int_0^\infty dt \int_{R\Sigma} \jpsi \cdot \bn d\sigma \ee
where $\bx_e^R$ is the position at which the trajectory first  crosses the
sphere $\PKR$  and $\IP^\psi$ is the quantum equilibrium measure, given by
the  density $|\psi|^2$.  This provides a natural definition of the
cross section measure. \end{rmrem}

\section*{Acknowledgments}
This work was supported in part  by
the DFG, by NSF Grant  No. DMS-9504556, and by the INFN.

\end{document}